\documentstyle[preprint,floats,prd,aps]{revtex}
\begin{document}
\preprint{\vbox {\hbox{RIKEN-AF-NP-274}}}

\draft
\title{Non-spectator contribution: A mechanism for inclusive $B\to X_s \eta'$ and exclusive $B\to K^{(*)}\eta'$ decays} \author{Mohammad R. Ahmady$^a$
\footnote{Email: ahmady@riken.go.jp}
, Emi Kou$^b$\footnote{Email: kou@fs.cc.ocha.ac.jp} and Akio
Sugamoto$^b$\footnote{Email:sugamoto@phys.ocha.ac.jp}}
\address{
$^a$LINAC Laboratory, The Institute of Physical
and Chemical Research (RIKEN)\\ 2-1 Hirosawa, Wako, Saitama 351-01,
Japan \\
$^b$ Department of Physics, Ochanomizu University \\
1-1 Otsuka 2, Bunkyo-ku,Tokyo 112, Japan}

\date{October 1997}
\maketitle
\begin{abstract}
We propose a mechanism which can explain the large $BR(B\to X_s\eta' )\approx 10^{-3}$ observed by CLEO.  In this mechanism, $\eta'$ is produced by the fusion of two gluons, one from QCD penguin $b\to sg$ and the other one emitted by the light quark inside the B meson.  The inclusive decay rate which is calculated via using the factorization assumption,  can easily account for the observed branching ratio.  We also estimate the exclusive branching ratio $BR(B\to K\eta' )=7.0\times 10^{-5}$ which is in good agreement with the experimental data and present our prediction for $K^*$ mode $BR(B\to K^*\eta' )=3.4\times 10^{-5}$.  
\end{abstract}
%
\newpage
\section{Introduction}
The CLEO collaboration has recently discovered an unexpectedly large branching ratio for the semi-inclusive hadronic $B\to X_s\eta'$ decay\cite{cleo}:
\begin{equation}
BR(B\to X_s\eta' )=(7.5\pm 1.5\pm 1.1)\times 10^{-4}\;\;\;2.0\le p_{\eta'}\le 2.7\;\; GeV\;\;.
\end{equation}
The corresponding exclusive decay rate has also been measured:
\begin{equation}
BR(B\to K\eta' )=(7.8 ^{+2.7}_{-2.2}\pm 1.0)\times 10^{-5}\;\; .
\end{equation}
Possible mechanisms behind this large production of fast $\eta'$ meson have been discussed in recent papers\cite{as,hz,yc,ht,kp}.  For example, Atwood and Soni(AS)\cite{as} have suggested the subprocess $b\to sg^*\to s\eta' g$ ($g^*$ and $g$ are virtual and real gluons, respectively) as the main underlying mechanism.  For this purpose, the standard model QCD penguin is used in conjunction with gluon anomaly driven $g^*g\eta'$ vertex.  The form factor for this vertex $H(q^2,0,m^2_{\eta'})$, where $q$ is the 4-momentum of the off-shell gluon $g^*$, is approximated by the constant $H(0,0,m^2_{\eta'})$ which in turn is extracted from $J/\psi \to\eta' \gamma$ decay.  However, the approximation $H(q^2,0,m^2_{\eta'})\approx H(0,0,m^2_{\eta'})$ turns out to be problematic for two reasons: 1) As pointed out by Hou and Tseng\cite{ht}, there is an implicit factor of $\alpha_s$ in $H$ which should be running with $q^2$ and this would suppress AS's estimate by a factor $1/3$.  2) On the other hand, Kagan and Petrov\cite{kp} have indicated that the momentum dependence of the form factor could be quite significant with the leading behaviour of the form $m_{\eta'}^2/(q^2-m_{\eta'}^2)$.  As a result, including this effect further reduces AS's result to about an order of magnitude below the observed branching ratio.  Cosequently, it has been suggested that the remedy could be in invoking new physics to increase $BR(b\to sg)$ to 10\%-15\% from its Standard Model value of nearly 0.2\%.

In this paper, we investigate the possibility that a somewhat different process might be the underlying mechanism for $B\to X_s\eta'$.  We propose a non-spectator process in which $\eta'$ is produced via fusion of the gluon from QCD penguin $b\to sg^*$ and another one emitted by the light quark inside B meson (Figure 1).  It is shown that a conservative estimate of the contribution of this mechanism can naturally account for the observed value.  We also calculate the branching ratios $BR(B\to K\eta' )$ and $BR(B\to K^*\eta' )$ in the context of factorization.  The latter turns out to be smaller than the former by a factor 2.

\section{Effective Hamiltonian}

The expression for Figure 1 is the product of three terms:\\
1) The effective neutral current flavor changing vertex $b\to sg$ which is as follows\cite{il}
\begin{equation}
A(b\to sg )=-i\lambda_t\frac{G_F}{\sqrt{2}}\frac{g_s}{8\pi^2}\left [E_0 \bar s(q^2g_{\mu\nu}-q_\mu q_\nu )\gamma^\nu (1-\gamma_5)T^ab-E'_0\bar s im_b\sigma_{\mu\nu}q^\nu (1+\gamma_5)T^ab\right ]\;\; ,
\end{equation}
where $\lambda_t=V_{tb}V^*_{ts}$ and $q$ is the gluon four-momentum.  In the process considered here, both chromo-electric and chromo-magnetic operators in (3) contribute.  However, since $\vert E_0\vert =\vert -4.92\vert \gg E'_0 =0.16$, the former operator is expected to have the dominant contribution.  It is argued that the inclusion of the chromo-magnetic operator would increase the inclusive decay rate by 20\% to 50\%\cite{as,ht}.  The purpose of this paper is to show that even a conservative estimate of our proposed mechanism can account for the observed branching ratio in an order of magnitude sense.  For this reason, in the rest of this work, we take into account only the dominant chromo-electric operator.\\
2) The gluon-gluon-psuedoscalar meson ($\eta'$ in this case) vertex which can be written as:
\begin{equation}
A^{\mu\sigma}(gg\to\eta')=H(q^2,p^2,m_{\eta'}^2)\delta^{ab}
\epsilon^{\mu\sigma\alpha\beta}q_\alpha p_\beta\;\; .
\end{equation}
$q$ and $p$ are four-momenta of the two gluons and $H$ is the relevant form factor which contains a factor of $\alpha_s$ implicitly.  AS made an estimate of $H(0,0,m_{\eta'}^2)\approx 1.8$ GeV$^{-1}$ using the decay mode $\psi\to\eta'\gamma$ which is expected to proceed mainly via on-shell gluons.  However, contrary to $H(q^2,p^2,m_{\eta'}^2)\approx H(0,0,m_{\eta'}^2)$ assumption utilized by AS, it is claimed that the momentum dependence of $H$ could be quite significant\cite{ht,kp} resulting in a suppression by an order of magnitude.  We show that the non-spectator mechanism suggested in this work can produce a large enough branching ratio which could match the observed value when such suppression factor is taken into account. \\
3) The emission of gluon by the light quark.  

By combining the above three terms one arrives at the effective Hamiltonian corresponding to Figure 1:
\begin{equation}
H_{eff}=CH(\bar s\gamma_\mu (1-\gamma_5)T^ab)(\bar q\gamma_\sigma T^a q)\frac{1}{p^2-M_g^2}\epsilon^{\mu\sigma\alpha\beta}q_\alpha p_\beta\;\; ,
\end{equation}
where
\begin{equation}
C=\lambda_t\frac{G_F}{\sqrt{2}}\frac{\alpha_s}{2\pi}E_0\;\; ,
\end{equation}
and the effective gluon mass $M_g$ regulates the singularity of the gluon propagator.  A re-arrangement of (5) via Fierz transformation:
\begin{eqnarray}
\nonumber
(\bar s\gamma_\mu (1-\gamma_5)T^ab)(\bar q\gamma_\sigma T^aq)=\frac{1}{9}
&&\left [(\bar s\gamma_\sigma\gamma_\rho\gamma_\mu (1-\gamma_5)q)(\bar q\gamma^\rho (1-\gamma_5)b)
+(\bar s\gamma_\sigma\gamma_\mu (1+\gamma_5)q)(\bar q(1-\gamma_5)b)\right . \\
&&\left . -\frac{1}{2}(\bar s\gamma_\sigma\sigma_{\rho\eta}\gamma_\mu q)(\bar q\sigma^{\rho\eta}(1-\gamma_5)b) +{\rm color\;\; octect}\right ]\;\; ,
\end{eqnarray}
simplifies the calculation of the hadronic matrix elements.  In fact, using the factorization assumption, only the first two terms in (7) contribute to $<\eta' X_s|H_{eff}|B>$.  Utilizing the definition of the B meson decay constant:
\begin{equation}
<0|\bar q\gamma^\mu\gamma_5b|B(p_B)>=f_Bp^\mu_B\;\; ,
\end{equation}
and its associated relation:
\begin{equation}
<0|\bar q\gamma_5b|B(p_B)>=-f_B\frac{M_B^2}{m_q+m_b}\;\; ,
\end{equation}
we obtain the relevant matrix element as follows:
\begin{equation}
<\eta' X_s|H_{eff}|B>=\frac{Cf_BH}{9(p^2-M_g^2)}\left [-(\bar s\gamma_\sigma\gamma_\rho\gamma_\mu (1-\gamma_5)q)p_B^\rho+\left (\frac{M_B^2}{m_q+m_b}\right )(\bar s\gamma_\sigma\gamma_\mu (1+\gamma_5)q)
\right ]\epsilon^{\mu\sigma\alpha\beta} q_\alpha p_\beta\;\; .
\end{equation}
Hereafter, we take the light quark mass $m_q=0$.
\section{recoil mass distribution and branching ratio}
The calculation of the differential decay rate is now straightforward, starting from the matrix element in eqn. (10).  We use the usual convention for the invariant variables $s={(p_{\eta'}+k' )}^2$, $t={(p_s+k' )}^2$ and $u={(p_s+p_{\eta'})}^2$.
\begin{eqnarray}
\nonumber\frac{d\Gamma (B\to X_s\eta' )}{dtdu}&=&\displaystyle \frac{C^2f_B^2H^2}{648\pi^3M_B^3{(p^2-M_g^2)}^2}\times \\
\nonumber\displaystyle
&&\left [p^2X\left \{ (W-Y-\frac{p^2}{2})(W-X)-XZ+(X-\frac{s+p^2}{2})W\right \} \right . \\
\nonumber\displaystyle &-&q^2ZW^2+XYZW+(s-2Y-q^2)(X-\frac{s+p^2}{2})W^2  \\
\nonumber\displaystyle &-&{\left (\frac{M_B^2}{m_b}\right )}^2\left \{ p^2(X-\frac{s+p^2}{2})(s-Y-q^2)-p^2q^2(W+Z-Y-\frac{p^2}{2}) \right . \\
\displaystyle &+& \left . \left . Y^2Z-(s-2Y-q^2)(W-Y-\frac{p^2}{2})Y\right \}
\right ]\;\; ,
\end{eqnarray}
where $W=(u-m_b^2)/2$, $X=(m_b^2-m_s^2+s)/2$, $Y=(m_{\eta'}^2-p^2-q^2)/2$ and $Z=(t-m_s^2)/2$.  $m_{X_s}^2=t$ is the invariant mass of the final state strange hadron.  The differential decay rate (11) depends on the virtualities of the internal gluons both explicitly and implicitly through the form factor $H$.  As we discussed in the previous section, $H(p^2,q^2,m_{\eta'}^2)$ is suppressed for large values of $p^2$ and $q^2$.  Therefore, the dominant contribution to the decay rate is expected to arise from small virtuality region.  On the other hand, in the non-spectator mechanism of Figure 1, due to kinematical freedom, one can impose a constraint such as $p^2=0$.  Consequently, $q^2$ can be expressed as:
\begin{equation}
q^2=m_b^2+m_{\eta'}^2-u+(t-m_s^2)\frac{u-m_b^2}{m_B^2-u}\;\; .
\end{equation}

Integrating eqn. (11) over $u$, we obtain the invariant mass distribution for the branching ratio, $dBR(B\to X_s\eta' )/dm_{X_s}$, depicted in Figure 2.  For our numerical evaluations, we have taken $m_b=4.5$ GeV, $m_s=0.15$\footnote{For phase space calculation, $m_s$ is taken to be the constituent quark mass $m_s^{\rm constituent}\approx 0.45$ GeV.}, $M_g\approx \Lambda_{QCD}\approx 0.3$ GeV, $\alpha_s=0.2$, $f_B=0.2$ GeV and $\vert V_t\vert=\vert V_{tb}V_{ts}^*\vert\approx\vert V_{cb}\vert\approx 0.04$ .  In order to obtain the total branching ratio with the experimental cut \mbox{$2.0\leq p_{\eta'}\leq 2.7$ GeV}, the differential decay rate (Figure 2) is integrated over the range \mbox{$m_s^{\rm constituent}=0.45\leq m_{X_s}\leq 2.32$ GeV} resulting in:
\begin{equation}
BR(B\to X_s\eta' )=4.7\times 10^{-3}\;\;\; 2.0\leq p_{\eta'}\leq 2.7\;\;{\rm GeV}\;\; .
\end{equation}
The momentum dependence of $H(q^2,0,m_{\eta'}^2)$ has not been taken into account in the above estimate.  However, even if one considers up to an order of magnitude suppression due to this form factor, our result indicates that the non-spectator mechanism is indeed the dominant process making up the bulk of the experimental data (1).
\section{Exclusive decays $B\to K\eta'$ and $B\to K^*\eta'$}
Using eqn. (10) in conjunction with the factorization assumption, one can write the hadronic matrix element for $B\to K\eta'$ as follows:
\begin{eqnarray}
<\eta' K|H_{eff}|B>=\displaystyle\frac{Cf_BH}{9(p^2-M_g^2)}\left [-<K|\bar s\gamma_\sigma\gamma_\rho\gamma_\mu (1-\gamma_5)q|0>p_B^\rho\right . \\
\nonumber\displaystyle\left . +\left (\frac{M_B^2}{m_b}\right )<K|\bar s\gamma_\sigma\gamma_\mu (1+\gamma_5)q|0>
\right ]\epsilon^{\mu\sigma\alpha\beta} q_\alpha p_\beta\;\; .
\end{eqnarray}
In fact, the second term in (14) does not contribute and the first term can be related to K-meson decay constant via the definition
\begin{equation}
<K(p_K)|\bar s\gamma^\mu\gamma_5 q|0>=f_Kp^\mu_K\;\; ,
\end{equation}
along with the identity
\begin{equation}
\gamma_\sigma\gamma_\rho\gamma_\mu =i\epsilon_{\sigma\rho\mu\nu}\gamma_5\gamma^\nu +g_{\sigma\rho}\gamma_\mu-g_{\sigma\mu}\gamma_\rho+g_{\rho\mu}\gamma_\sigma\;\; .
\end{equation}
As a result, the matrix element (14) can be simplified to:
\begin{equation}
<\eta' K|H_{eff}|B>=-i\frac{CHf_Bf_K}{9(p^2-M_g^2)}\left (p_B.qp_K.p-p_B.pp_K.q\right )\;\; ,
\end{equation}
leading to the exclusive decay rate:
\begin{equation}
\Gamma (B\to K\eta' )=\frac{C^2H^2f_B^2f_K^2}{1944\pi M_g^4}{\vert \vec p_K\vert}^3\left (m_{\eta'}^2+4{\vert \vec p_K\vert}^2\right )p_0^2\;\; ,
\end{equation}
which is derived by imposing the $p^2=0$ constraint.  $\vert\vec p_K\vert$ and $p_0$ are the three momentum of the K meson and the energy change of the light quark in B meson rest frame,  respectively:
\begin{eqnarray}
\vert \vec p_K\vert &=&{\left [\frac{{(m_B^2+m_K^2-m_{\eta'}^2)}^2}{4m_B^2}-m_K^2\right ]}^{\frac{1}{2}}\;\; ,\\
p_0&=&\frac{m_B^2-m_b^2}{2m_B}-E_q\;\; ,
\end{eqnarray}
where $E_q$ is the energy of the light quark in K meson.  To proceed with the numerical evaluation of $BR(B\to K\eta' )$, one can assume an appropriate model to estimate $E_q$.  However, roughly speaking, one would expect $E_q=m_K-m_s^{\rm constituent}\approx 0.05$ GeV.  The exclusive branching ratio is then estimated to be
\begin{equation}
BR(B\to K\eta' )=7.0\times 10^{-5}\;\; ,
\end{equation}
which is in good agreement with experimental data (2).

In the same manner, the matrix element relevant for $B\to K^*\eta'$ decay can be obtained from (10):
\begin{eqnarray}
<\eta' K^*|H_{eff}|B>=\displaystyle -i\frac{CHf_Bf_{K^*}}{9(p^2-M_g^2)}\left [p_B.q\epsilon .p-p_B.p\epsilon .q \right . \\
\nonumber\displaystyle\left . +\frac{2m_B^2m_s}{m_bm_{K^*}^2}\left (-i\epsilon^{\mu\sigma\alpha\beta}{p_{K^*}}_\mu\epsilon_\sigma q_\alpha p_\beta +p_{K^*}.q\epsilon .p-p_{K^*}.p\epsilon .q\right )
\right ]\;\; ,
\end{eqnarray}
where the polarization vector $\epsilon$ and the decay constant $f_{K^*}$
of $K^*$ appear in (21) via the definition:
\begin{equation}
<K^*(p_{K^*},\epsilon )|\bar s\gamma^\mu q|0>=f_{K^*}\epsilon^\mu \;\; ,
\end{equation}
and its follow up:
\begin{equation}
<K^*(p_{K^*},\epsilon )|\bar s\sigma^{\mu\nu} q|0>=i\frac{m_s}{m_{K^*}^2}f_{K^*}(p^\mu_{K^*}\epsilon^\nu -p^\nu_{K^*}\epsilon^\mu ) \;\; .
\end{equation}
Eqn. (22) in conjunction with the constraint $p^2=0$ leads to the following expression for the exclusive decay rate:
\begin{eqnarray}
\nonumber\Gamma (B\to K^*\eta' )&=&\frac{C^2H^2f_B^2f_{K^*}^2}{1296\pi M_g^4m_B^2}\vert\vec p_{K^*}\vert \int^1_{-1}F(x)dx\;\;, \\
\nonumber
F(x)&=&{(C_1+C_3C_5)}^2\frac{C_4^2}{m^2_{K^*}}+{(C_2+C_4C_5)}^2(-q^2+\frac{C_3^2}{m_{K^*}^2}) \\
&-&2(C_1+C_3C_5)(C_2+C_4C_5)(-C_6+\frac{C_4C_3}{m_{K^*}^2}) \\
\nonumber
&+&C_5^2\left [m_{K^*}^2C_6^2-C_3C_6C_4-C_4(C_3C_6-q^2C_4)\right ]\;\; '
\end{eqnarray}
where
\begin{eqnarray}
\nonumber C_1&=&\frac{m_B^2+m_{\eta'}^2-m_{K^*}^2}{2}-C_2\;\; ,\\
\nonumber C_2&=&m_Bp_0\;\; , \\
\nonumber C_3&=&\frac{m_B^2-m_{\eta'}^2-m_{K^*}^2}{2}-C_4\;\; , \\
\nonumber C_4&=&\left [{(m_{K^*}^2+{\vert\vec p_{K^*}\vert}^2)}^{1/2}-\vert\vec p_{K^*}\vert x\right ]p_0\;\; , \\
\nonumber C_5&=&\frac{2m_sm_B^2}{m_bm_{K^*}^2}\;\; , \\
\nonumber C_6&=&\frac{m_{\eta'}^2-q^2}{2}\;\; , \\
\nonumber \vert\vec p_{K^*}\vert &=&\frac{{[(m_B^2-{(m_{\eta'}+m_{K^*})}^2)(m_B^2-{(m_{\eta'}-m_{K^*})}^2)]}^{1/2}}{2m_B}\;\; .
\end{eqnarray}
$p_0$ is the same as in eqn. (20) and $q^2$ is obtained from eqn. (12) by substituting $m_{K^*}^2$ and $m_B^2-2m_BE_q$ for $t$ and $u$, respectively.  In order to estimate the branching ratio for $B\to K^*\eta' $, in analogy with $B\to K\eta'$ exclusive decay, we take the energy of the light quark in $K^*$ to be:
$$E_q\approx m_{K^*}-m_s^{constituent}\approx 0.44 {\rm GeV}\;\; ,$$
 which results in\footnote{We estimate $f_{K^*}\approx 205$ GeV$^2$ by using the experimental value of the ratio $\Gamma (\tau\to {K^*}^-\nu_\tau )/\Gamma (\tau\to K^-\nu_\tau )$ and $f_K=0.167$ GeV \cite{pdg}.}:
\begin{equation}
BR(B\to K^*\eta' )=3.4\times 10^{-5}\;\; .
\end{equation}
We note that the results for exclusive decays should not be altered significantly due to momentum dependence of $H$.  This is due to the fact that, unlike the inclusive process, $q^2$ for these decays is fixed at around $1-3$ GeV$^2$.  Measurement of $K^*$ mode will be a crucial testing ground for various mechanisms suggested for $\eta'$ production in hadronic B decays.  For example, our prediction is in contrast to $\Gamma (B\to K^*\eta' )\approx 2\Gamma (B\to K\eta' )$ obtained from the proposed $b\to c\bar cs\to\eta' s$ process\cite{hz}. 
\section{Summary}
We calculated a non-spectator contribution to the inclusive B meson decay into $\eta'$ and hadrons containing a strange quark.  The result indicates that this mechanism could explain the large experimental branching ratio $BR(B\to X_s\eta' )$ obtained by CLEO.  Our estimated exclusive branching ratio $BR(B\to K\eta')$ agrees with experiment as well.  The experimental confirmation of the predicted branching ratio for $B\to K^*\eta'$ will give a strong support to the suggestion that the non-spectator mechanism is indeed the underlying process for the above decay modes.   

\section{acknowledgement}
The authors thank Professor I. A. Sanda for suggesting this project and useful discussions.  M. A. acknowledges support from the Science and Technology Agency of Japan under an STA fellowship.

\newpage
{\center \bf \huge Figure Captions}
\vskip 3.0cm
\noindent
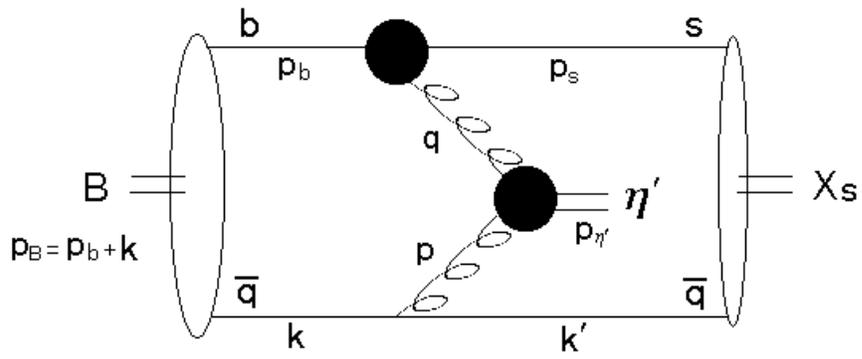
{\bf Figure 1}: Non-spectator contribution to $B\to\eta' X_s$. \\
\vskip 0.5cm
\noindent
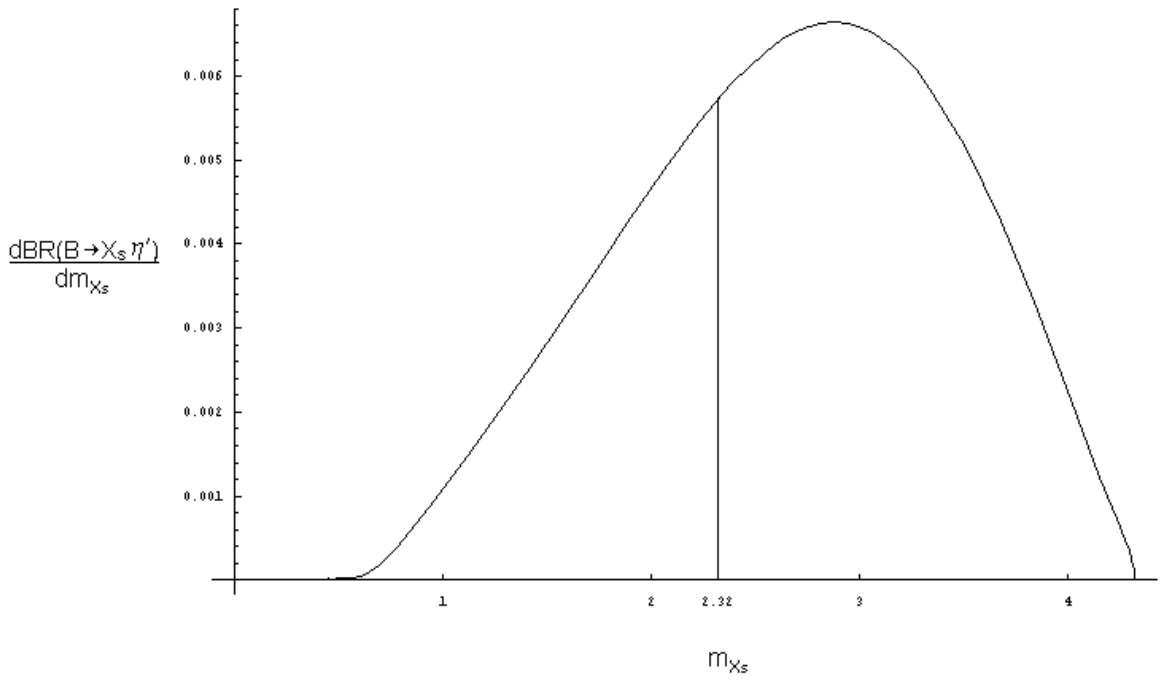
{\bf Figure 2}: The recoil mass distribution of the branching ratio for the  inclusive decay $B\to\eta' X_s$. \\
\vskip 0.5cm

\end{document}